\documentclass[12pt]{article} 
\usepackage{epsfig}

\def\half{{\textstyle{\frac{1}{2}}}}
\def\eq#1{eq.~(\ref{#1})}
\def\eqs#1#2{eqs.~(\ref{#1})\ and (\ref{#2})}
\def\ie{\hbox{\it i.e.}}

\def\eg{\hbox{\it e.g.}}

\def\bra#1{\left\langle #1\right|}
\def\ket#1{\left| #1\right\rangle}
\def\braket#1#2{\VEV{#1 | #2}}
\def\VEV#1{\left\langle #1\right\rangle}

\def\gesim{\,{\raise-3pt\hbox{$\sim$}}\!\!\!\!\!{\raise2pt\hbox{$>$}}\,}
\def\lesim{\,{\raise-3pt\hbox{$\sim$}}\!\!\!\!\!{\raise2pt\hbox{$<$}}\,}
\def\boldoverdot{\,{\raise6pt\hbox{\bf.}\!\!\!\!\>}}

\def\ie{{\it i.e.}}

\def\acal{{\cal A}}
\def\bcal{{\cal B}}

\def\dcal{{\cal D}}

\def\ical{{\cal I}}

\def\lcal{{\cal L}}

\def\tcal{{\cal T}}

\def\diag{\hbox{\diag}}

\def\doubleundertext#1{
{\undertext{\vphantom{y}#1}}\par\nobreak\vskip-\the\baselineskip\vskip4pt%
\undertext{\hbox to 2in{}}}
\def\inbox#1{\vbox{\hrule\hbox{\vrule\kern5pt
     \vbox{\kern5pt#1\kern5pt}\kern5pt\vrule}\hrule}}
\def\sqr#1#2{{\vcenter{\hrule height.#2pt
      \hbox{\vrule width.#2pt height#1pt \kern#1pt
         \vrule width.#2pt}
      \hrule height.#2pt}}}

\def\today{\ifcase\month\or
  January\or February\or March\or April\or May\or June\or
  July\or August\or September\or October\or November\or December\fi
  \space\number\day, \number\year}
\def\pmb#1{\setbox0=\hbox{#1}%
  \kern-.025em\copy0\kern-\wd0
  \kern.05em\copy0\kern-\wd0
  \kern-.025em\raise.0433em\box0 }
\def\sumprime_#1{\setbox0=\hbox{$\scriptstyle{#1}$}
  \setbox2=\hbox{$\displaystyle{\sum}$}
  \setbox4=\hbox{${}'\mathsurround=0pt$}
  \dimen0=.5\wd0 \advance\dimen0 by-.5\wd2
  \ifdim\dimen0>0pt
  \ifdim\dimen0>\wd4 \kern\wd4 \else\kern\dimen0\fi\fi
\mathop{{\sum}'}_{\kern-\wd4 #1}}
\advance \voffset -.75 in
\advance \hoffset -.5 in

\def\sgn{{\,\rm Sgn}}
\newcommand{\nc}{\newcommand}
\nc{\beq}{\begin{equation}}  \nc{\eeq}{\end{equation}}
\nc{\bea}{\begin{eqnarray}}  \nc{\eea}{\end{eqnarray}}
\nc{\baa}{\begin{array}}     \nc{\eaa}{\end{array}}
\nc{\bit}{\begin{itemize}}   \nc{\eit}{\end{itemize}}
\nc{\ben}{\begin{enumerate}} \nc{\een}{\end{enumerate}}
\nc{\bce}{\begin{center}}    \nc{\ece}{\end{center}}

\begin{document}

\begin{titlepage}

\title {{\vskip-1in \hfill{\small MCTP-03-24}\newline
\rightline{\small hep-th/0305056} } \vskip.5in
Squeezed States in the de~Sitter Vacuum}
\author{Martin B. Einhorn\footnote{\texttt{meinhorn@umich.edu}}\ \ and 
Finn Larsen\footnote{\texttt{larsenf@umich.edu}} \\
\centerline{\it\normalsize Michigan Center for Theoretical Physics,
Randall Laboratory of Physics} \\  
\centerline{\it\normalsize The University of Michigan, Ann Arbor, MI 48109-1120} }

\date{}

\maketitle

\begin{abstract}
We discuss the treatment of squeezed states as excitations in the Euclidean
vacuum of de~Sitter space. A comparison with the treatment of these states as candidate
no-particle states, or alpha-vacua, shows important differences already in the free theory. At
the interacting level alpha-vacua are inconsistent, but squeezed state excitations seem
perfectly acceptable. Indeed, matrix elements can be renormalized in the excited states
using precisely the standard local counterterms of the Euclidean vacuum. Implications for
inflationary scenarios in cosmology are discussed.
\end{abstract}

\thispagestyle{empty}
\end{titlepage}

\setcounter{page}{1}

\section{Introduction}
Some years ago it was understood that there exists a two parameter set of de~Sitter
invariant states that may be regarded as candidate no-particle states, or vacua, for a
scalar field in de~Sitter space~\cite{Chernikov:zm,Mottola,Allen}. These so-called 
alpha-vacua are alternatives to the Euclidean or Bunch-Davies vacuum~\cite{Bunch:yq} usually
considered appropriate for cosmological inflation. Recently there has been much work on
the interpretation of these alpha-vacua as ambiguities in the low energy theory
parametrizing physics beyond the Planck scale~\cite{udan,transplanck}. Alpha-vacua are
also important in considerations regarding de~Sitter holography~\cite{Strominger,SV}. 

In a previous paper~\cite{Einhorn:2002nu} (hereinafter referred to as {\bf I}), we argued
that the choice of de~Sitter invariant vacuum is in fact ambiguious only for free field
theory, since the Feynman rules for interacting fields lead to ill-defined  loop diagrams
for all but the Euclidean vacuum. Similarly objections have been raised to the nonlocal
counterterms that alpha-vacua would require~\cite{Banks}. Against these concerns stands
the fact that alpha-vacua clearly resemble squeezed states in quantum
optics~\cite{optics,Barnett}; so the question naturally arises as to whether one could not
regard such states as excited states in the Euclidean vacuum. Just as in quantum optics,
it must be that the problems encountered in {\bf I} can be avoided. Moreover, from this point
of view one does not expect non-local counterterms, since transition amplitudes involving
excited states should be rendered finite by the same local counterterms that renormalize
vacuum amplitudes.  

The purpose of this paper is to resolve the obvious tension between these various results
and expectations. The first step towards this goal will be to discuss some key
differences, present already in the free theory, between the treatment of squeezed state
excitations of the Euclidean vacuum and the interpretation of these states as vacua in
their own right. For example, we will argue that the time-ordered two point correlators
are in fact different in these two situations. Another (related) difference is that the
notorious antipodal singularities of the two point correlators are associated with
sources, when the state is treated as an excitation, but not when it is interpreted as a
non-standard vacuum. These features of the free theory lead to the suspicion that squeezed
states might be perfectly viable as excited states in the Euclidean vacuum, even if they
are unacceptable as vacua.\footnote{While this work was underway, a
paper appeared~\cite{Goldstein:2003ut} which makes the same distinction as we, and also
reaches the conclusion that excited states are viable. Unfortunately, we disagree on both
the Feynman rules and the renormalization counterterms.  In fact, the prescription for
loop diagrams given in~\cite{Goldstein:2003ut} seems to be precisely the one criticized in
{\bf I} as being mathematically nonsensical.}  

The true test of these ideas is the full interacting theory. In section 4 we explain how
the Feynman rules can be formulated for squeezed states, treated as excited states in the
Euclidean vacuum. According to these computational rules, the divergences will be those of
the Euclidean vacuum, removable by the standard local counterterms, as expected. Our
treatment could easily be extended to other excited states, such as to coherent states.
It is interesting that, for consistency reasons alone, one can rule out the interpretation
of the alpha-vacua as no-particle states, yet admit such states as excited states of the
unique Euclidean vacuum. Our results illustrate some of the powerful constraints resulting
from going beyond free field theory to consider interacting fields.

The outline of the paper is as follows:  in the next section, we give a short review of
squeezed states and alpha-vacua.  In section 3 we consider the free correlators in the two
situations and, in section 4, we proceed to the interacting theory and discuss Feynman
rules and renormalization. Finally, in section 5, we conclude by discussing implications
our result for the Hadamard condition of quantum field theory in curved space, and for
cosmology. 

\section{On Vacua and Squeezed States}
\label{sect:squeezed}

Let us begin by recalling the notion of a vacuum in curved space. We adopt the notation of
{\bf I}, mostly in common with~\cite{Allen}. The quantum field  for a free scalar is
written in terms of a mode expansion
\beq\label{modeexp}
\phi(x) = \sum_{n} [ a_{n} u_{n}(x) + a_{n}^{\dag}u_{n}^{*}(x)]~,
\eeq
and the corresponding specification of the ``no-particle" or ``vacuum" state
\beq
\label{vacdef}
a_{n}|{\rm vac}\rangle = 0 ~.
\eeq
As usual, the subscript $n$ denotes all quantum numbers on which the mode depends;Fock
states are built up by applications of the $a_n^\dag$, and the Hilbert space is defined by
their completion. The standard choice of the functions \{$u_{n}$\} is variously known as
the Euclidean vacuum or the Bunch-Davies vacuum. Numerous arguments in favor of the
standard choice are given in the literature, {\it e.g.}~\cite{Birrell,Fulling:nb}. Since we already
reviewed these arguments in {\bf I} we shall not do so again, apart from noting that the
Euclidean vacuum also is the vacuum singled out by the considerations of {\bf I} that
interactions be introduced consistently. 

Alternate definitions of vacua are associated with different choices for the mode
functions.  Since each choice is a complete set of modes, new choices may be expressed in
terms of the Euclidean modes \{$u_{n}$\} as a Bogoliubov transformation which, for our
purposes, can be taken of the form  
\beq\label{bog}
\widetilde {u_{n}}(x) = 
 u_n(x)\cosh\alpha_n + u_n^*(x)e^{i\beta_n} \sinh\alpha_n~.
\eeq
The corresponding field may be expanded as 
\beq
\label{modetilde}
\phi(x) = \sum_{n} [ \widetilde{a_{n}} \widetilde{u_{n}}(x) +
\widetilde{a}_{n}^{\dag}\widetilde{u}_{n}^{*}(x)]~,
\eeq
where the associated Bogoliubov transform of the operators is 
\beq\label{atilde}
\widetilde{a_n}=  a_n\cosh\alpha_n - a_n^\dag e^{-i\beta_n} \sinh\alpha_n~,
\eeq
with corresponding state
\beq\label{alphavac}
\widetilde{a_{n}}\ket{\underline{\alpha},\underline{\beta} } = 0 ~,
\eeq
for all $n.$

In order to interpret the candidate vacuum states as excited states note that 
the Bogoliubov transformation \eq{atilde} may be implemented by a 
unitary transformation~\cite{optics,Barnett}
\beq\label{similarity}
S(\underline{\xi}) a_n S(\underline{\xi})^\dag = \widetilde{a_n}~,
\eeq
where $\xi_n\equiv e^{-i\beta_n}\alpha_n$ and
\beq\label{squeeze}
S(\underline{\xi}) \equiv 
\exp \sum_n \half\left[ \xi_n a_n^{\dag 2}-\xi_n^* a_n^2 \right]~,
\eeq
The operator $S$ can be rewritten as~\cite{Barnett} 
\bea\label{normal}
S(\underline{\xi}) &=&\exp \sum_{n}\left[ \half a_n^{\dag 2} 
e^{-i\beta_n}\tanh\alpha_n\right] 
\exp \sum_n \left[ -\half\left(a_na_n^\dag + a_n^\dag a_n\right) 
\ln(\cosh\alpha_n)\right]\nonumber\\
&\times& \exp \sum_n \left[-\half a_n^2e^{i\beta_n}\tanh\alpha_n\right]~,
 \eea
so that the state \eq{alphavac} may be represented in terms of the original
quanta as
\beq\label{alphastate}
\ket{\underline{\alpha},\underline{\beta} }=S(\underline{\xi})\ket{0}=
\exp \sum_n \left[-\half\ln(\cosh\alpha_n)\right]
\exp\sum_n\left[\half\tanh \alpha_n e^{-i\beta_n} a_n^{\dag 2}\right]\ket{0}~.
\eeq
This formula is of central importance, since it shows that the vacuum defined by
(\ref{alphavac}) can in general be represented as a state in the Euclidean theory.

One of the nice properties of the Euclidean modes \{$u_{n}$\} is that they respect the de
Sitter invariance of the background. The alternate vacua \{$\tilde{u}_{n}$\} break this
symmetry except when $\alpha_n\equiv\alpha$ and $\beta_n\equiv\beta$ are independent of
mode-number $n$. The family of de Sitter invariant vacua parametrized by $\alpha,\beta$
are those found by Mottola~\cite{Mottola} and Allen~\cite{Allen}. In most of our
considerations we will keep the general $n$-dependence and refer to alpha-vacua, when we
interpret these states as no-particle states of the system, and alpha-states (or sometimes
squeezed states) when we treat them as excited states in the Euclidean vacuum. The
de~Sitter invariant alpha-vacua, independent of $n$, are the MA-vacua.

The discussion so far tacitly assumes that the mode numbers are discrete, which is true
for certain coordinate systems.  However, for the planar coordinates typical of FRW
models, the notation must be refined in order to deal with continuous indices.  One must
replace the discrete index ${n}$ by the wave number $\vec{k}$, introduce wave packets in
momentum space, and write the Bogoliubov transformation \eq{atilde} in the non-local form
\beq
\widetilde{a_{\,\vec k}}=  a_{\,\vec k}\cosh\alpha_{\vec k} - a_{-\vec k}^\dag 
e^{-i\beta_{\vec k}} \sinh\alpha_{\vec k}~.
\eeq
As is familiar from Minkowski space, the treatment becomes more cumbersome in order to
deal with the mathematics of distributions rather than functions.  To keep formulae as
simple as possible, we will retain the discrete notation with the understanding that it
can be adapted to the continuous case as necessary.

It can be shown that a non-zero phase $\beta_n$ is associated with CPT
violation~\cite{Allen} so, for most applications, it is presumed that $\beta_n=0.$  To
simplify the formulae here, we will also make that assumption, although it would not be
difficult to extend our treatment to the general case.  Accordingly, $\xi_n=\alpha_n,$ and we
will abbreviate the state $\ket{\underline{\alpha},\underline{0}}$ as
$\ket{\underline{\alpha}}.$ 

The overlap between the alpha-state (\ref{alphastate}) and the Euclidean vacuum is 
\beq\label{overlap}
\braket{0}{\underline{\alpha}}=\prod_n \frac{1}{\sqrt{\cosh{\alpha_n}}}~.
\eeq
The $\alpha_n=\alpha$ are $n$-independent for the MA-vacua; so the overlap vanishes,  as
does the overlap with any Fock state of the Euclidean theory. Formally this shows that these states
$\ket{\underline{\alpha}}$ are orthogonal to the Euclidean Hilbert space.  If taken at
face value, this means the  $\ket{\underline{\alpha}}$ cannot be described as excited
states, but rather must be treated as vacua. Then there would be an orthogonal Hilbert
space for each $\alpha$, with the no-particle state annihilated by the appropriate
$\widetilde{a_n}$ and the Fock states built by application of the corresponding
$\widetilde{a_n}^\dag$.

However, in our view, this formal argument fails to represent a sensible approximation to
the physics. According to the modern view of renormalization~\cite{Weinberg} every local
field theory should be regarded as an effective field theory below some high energy scale,
the cut-off. As a result, the alpha-states, whether dependent on $n$ or not, should be
treated as having finite overlap with states in the Euclidean theory and hence can be regarded as
excited states. In fact, because of difficulties with defining the Feynman rules for 
MA-vacua, these do not really represent alternatives  to the Hilbert space based on the
Euclidean vacuum. 

\section{Correlators in Free Field Theory}
\label{sect:free}
We now proceed to explain how the treatment of excited states differ from that of vacua,
by discussing the correlators of the free theory.
Using \eq{alphastate}, Wightman functions for correlators in the states
$\ket{\underline{\alpha}}$ may be represented in terms of operators in the Euclidean
vacuum as 
\beq\label{phitilde}
W_\alpha(x,y)=\bra{\underline{\alpha}}\phi(x)\phi(y)\ket{\underline{\alpha}}=
\bra{0}\widetilde{\phi(x)}\widetilde{\phi(y)}\ket{0},
~~{\rm where}~~\widetilde{\phi(x)}\equiv
S^\dag(\underline{\alpha})\phi(x)S(\underline{\alpha})~.
\eeq
To evaluate this further we use \eq{similarity} to write
\bea\label{phitilde2}
\widetilde{\phi(x)}&=&
\sum_n \cosh\alpha_n \left[a_n u_n(x) + a_n^\dag u_n^*(x) \right] +
\sum_n \sinh\alpha_n \left[a_n u_n(\overline{x}) + a_n^\dag
u_n^*(\overline{x}) \right],\\
\label{abfields} &\equiv& \acal_\alpha(x) + \bcal_\alpha(\overline{x})~.
\eea
where we have adopted the choice of  basis introduced by Allen~\cite{Allen} for which
$u_n^*(x)=u_n(\overline{x}),$ where $\overline{x}$ represents the antipode of $x$.  

In the MA-vacua, for which $\alpha_n=\alpha$ is independent of $n$, \eq{phitilde2} can be
written formally as 
\beq\label{phitilde3}
\widetilde{\phi(x)}= \phi(x) \cosh\alpha +  
\phi(\overline{x})\sinh\alpha~,
\eeq
and so the Wightman function \eq{phitilde} becomes
\bea\label{wightman}
W_\alpha(x,y)\!&=&\! W_0(x,y)\cosh^2\alpha+
W_0(\overline{x},\overline{y})\sinh^2\alpha +
\half\left[W_0(x,\overline{y})+
W_0(\overline{x},y)\right]\sinh2\alpha~,\nonumber\\
&=&\! W_0(x,y)\cosh^2\alpha+
W_0^*(x,y)\sinh^2\alpha +
\half\left[W_0(x,\overline{y})+
W_0^*(x,\overline{y}\right]\sinh2\alpha~,
\eea
where, in the second line, we used $W_0(x,y)=W_0^*(y,x)$ and
$W_0(\overline{x},\overline{y})=W_0(y,x)$ in the Allen basis. This equation is central for
our interpretation of alpha-vacua: if we treat  $\ket{\underline{\alpha}}$ as a true
vacuum, the correlator $W_\alpha(x,y)$ is simply the amplitude for creation of a particle
at $y$ and annihilation at $x$. In contrast, if we treat the same formula as a statement
in the Euclidean vacuum we see that the amplitude has components not only involving creation 
of a particle at $y$ followed by its annihilation at $x,$ but also involving creation of a 
particle at $\overline{y}$ followed by its annihilation at either $\overline{x}$ or $x.$
The apparent non-local  and acausal creation and annihilation of particles is perhaps unfamiliar; 
however, it is not paradoxical in view of the fact that we are postulating a highly correlated background
state.  

The significance of this interpretation becomes clear when we consider time-ordered
correlators. Allen~\cite{Allen} treats the $\ket{\underline{\alpha}}$ as true vacua and
defines
\beq\label{MAprop}
iG^F_\alpha(x,y)\equiv \bra{\underline{\alpha}}\tcal\Big(\phi(x)\phi(y)\Big)
\ket{\underline{\alpha}}=
\Theta(x,y)W_{\alpha}(x,y)+\Theta(y,x)W_{\alpha}(y,x),
\eeq
where the time-ordering symbol is $\Theta(x,y) \equiv (1+{\rm Sgn}(x,y))/2,$ 
with ${\rm Sgn}(x,y)\equiv0$ if $x$ and $y$
are spacelike separated, while for timelike or lightlike separations, 
${\rm Sgn}(x,y)\equiv+1$ if $x>y$, or $\equiv-1$ if $x<y$. This expression can
be written
\beq
G^F_\alpha(Z)= \cosh^{2}\alpha\, G_0^{F}(Z)+ 
\sinh^{2}\alpha (G_0^F (Z))^{*} + 
\half\sinh 2\alpha\left(G_0^{F}(-Z) + {\rm c.c.}\right)~.
\eeq
When we treat $\ket{\underline{\alpha}}$ as an excited state it is more natural to
introduce time-ordering according to the definition
\beq\label{npoint}
\tcal\bra{\underline{\alpha}}\phi(x_1)\phi(x_2)\ldots
\phi(x_n)\ket{\underline{\alpha}}\equiv
\tcal\bra{0}\widetilde{\phi(x_1)}\widetilde{\phi(x_2)}\ldots
\widetilde{\phi(x_n)}\ket{0}~.
\eeq
The meaning of the right hand side of this expression is that the fields
$\widetilde{\phi(x)}$ should be expressed first as the linear combinations \eq{abfields},
and then time-ordering is carried out with respect to the arguments of the fields
$\acal(x)$ and $\bcal(\overline{x})$. In the case of the MA-states, this reduces to linear
combinations of the field $\phi$ itself, \eq{phitilde3}.

That \eq{npoint} is the correct definition of time-ordering when we treat
$\ket{\underline{\alpha}}$ as an excited state follows directly from the physical
interpretation of \eqs{abfields}{phitilde3}. In the context of the Fock space of the
Euclidean theory, the first sum $\acal_\alpha(x)$ in \eq{abfields} involves the creation
and annihilation of particles at the point $x$, while the second sum
$\bcal_\alpha(\overline{x})$ must be interpreted as the creation and annihilation of
particles at the antipodal point $\overline{x}$. 

Alternatively we can derive \eq{npoint} from the Feynman path integral (FPI)
representation for the generating functional 
\beq\label{fpi}
\exp\{iW[J]\}= \int\dcal\phi \exp\left[{iS[\phi]+i\int dx
\sqrt{g}J(x)\phi(x)}\right]~.
\eeq
For the Euclidean vacuum, this formal expression may be properly defined by Wick rotation
from Euclidean signature, just as is normally done in Minkowski background. Accordingly,
it is the generating functional for time-ordered Green's functions in the unique Euclidean
vacuum. For the alpha-states, it is clear that the corresponding generating functional
should be taken as 
\beq\label{fpi2}
\exp\{i\widetilde{W_\alpha}[{J}]\}= \int\dcal\phi \exp\left[{iS[\phi]+i\int dx
\sqrt{g}{J(x)}\widetilde{\phi(x)}}\right].
\eeq
Since the products of fields are automatically time-ordered with respect to their
arguments by the FPI, the Green's functions so generated will correspond to the
prescription given above for the right-hand side of \eq{npoint}.

Let us summarize. We have introduced two types of time-ordering: \eq{MAprop} for the
vacuum interpretation and  \eq{npoint} for the excited states. The crucial point simply is
that these, quite manifestly, are different
\beq\label{tprod}
\bra{\underline{\alpha}}\tcal\left(\phi(x)\phi(y)\right)
\ket{\underline{\alpha}} \ne 
\tcal\bra{\underline{\alpha}}\phi(x)\phi(y)\ket{\underline{\alpha}}~.
\eeq

To understand what the difference in time-ordering procedure means, recall that the
Feynman propagator of the MA-vacua, satisfies the same equation as the Euclidean
propagator, viz.,
\beq\label{kg1}
(\nabla_x^2+m^2) G^F_\alpha(x,y)=-\delta(x,y), {\rm where}\ 
\delta(x,y)\equiv\frac{\delta(x-y)}{\sqrt{g(x)}}~.
\eeq
Thus, as emphasized by Allen~\cite{Allen}, the difference between the MA-propagator and
the Euclidean propagator satisfies the {\it homogeneous} Klein-Gordon equation, even
though the MA-propagator is singular both for $x=y$ and $x=\overline{y}.$  It is this
peculiar singularity structure that leads to difficulties for interacting fields in the
MA-vacua~\cite{Einhorn:2002nu}.

In contrast, the two-point functions in the squeezed states is given by the right-hand-side 
of \eq{tprod} and may be denoted $i\widetilde{G^F_\alpha}$. This expression
corresponds to linear combinations of time-ordered products in the Euclidean vacuum in
exactly the same way as the ordinary products in \eq{wightman}. We therefore find
\beq\label{feynman1}
\widetilde{G^F_\alpha}(x,y) \!=\!
G^F_0(x,y)\cosh^2\alpha+G^F_0(\overline{x},\overline{y})\sinh^2\alpha +
\half\left[G^F_0(x,\overline{y})+ G^F_0(\overline{x},y)\right]\sinh2\alpha~.
\eeq
Using $G^F_0(x,y)=G^F_0(\overline{x},\overline{y})$, the four terms may combined to two,
\beq\label{feynman2}
\widetilde{G^F_\alpha}(x,y) \!=\! G^F_0(x,y)\cosh2\alpha +
G^F_0(x,\overline{y})\sinh2\alpha~.
\eeq
This implies that, in contrast to \eq{kg1}, 
\beq\label{kg2}
(\nabla_x^2+m^2) \widetilde{G^F_\alpha}(x,y)=
-\delta(x,y)\cosh2\alpha -
\delta(x,\overline{y})\sinh2\alpha~,
\eeq
so that there really is particle creation and annihilation at the antipodal point in
squeezed states. A source associated with the antipodal singularity is one of the
ingredients needed if one wants to treat this singularity as an image as, for example,
in~\cite{elliptic}.

Let us write our time ordered correlators in the conformally massless case where formulae
can be made explicit.  The propagator for the vacuum case
\beq
iG_F(Z) = {1\over 8\pi^2}\left[ {\cosh^2\alpha\over Z-1-
i\epsilon}+{\sinh^2\alpha\over Z-1+i\epsilon}-{\sinh 2\alpha\over Z+1}
\right]~,
\label{GFvac}
\eeq
where, in terms of the embedding coordinates $X(x)$ and $Y(y)$, $Z=-X\cdot Y$. Also recall
that, In the embedding coordinates, the antipode of $X$ is simply $-X$, accounting for the
appearance of $-Z$. In contrast, the time-ordered correlator for the excited state is 
\beq
iG_F(Z) = {1\over 8\pi^2}\left[ {\cosh 2\alpha\over Z-1-i\epsilon} - {\sinh
2\alpha\over Z+1+i\epsilon}
\right]~.
\label{GFexc}
\eeq
The singularity structure of \eq{GFvac} and \eq{GFexc} is completely different: for vacua
the $i\epsilon$ prescriptions are mixed, and the anti-podal singularity is simply the
principal value; for excited states a uniform $i\epsilon$ prescription is applied.

We should emphasize that we are {\it not} suggesting that one simply replace the MA-propagator 
in the $\alpha$-vacua with $\widetilde{G^F_\alpha},$ \eq{feynman1}.  We only
wish to indicate that time-ordered field correlators in the presence of background
squeezed states are different from the time-ordering of field operators in the 
$\alpha$-vacua. This opens the possibility that the physics of the two situations are 
different when one goes beyond free field theory.

Let us emphasize this point by considering other Green's functions. The MA-propagator
\eq{MAprop} can be written
\beq\label{green}
iG^F_\alpha(x,y)={1\over 2}\left[G_\alpha^{(1)}(x,y) +
i\sgn(x,y)D_\alpha(x,y)\right]~,
\eeq
where the symmetric term $G_\alpha^{(1)}$ is called the Hadamard function; and $D_\alpha,$
the commutator function.  The various two-point functions are given in terms of their
Euclidean counter-parts by~\cite{Allen} 
\bea\label{newgfcts}
G^{(1)}_{\alpha}(x,y) \!\!\! &= &\!\!\! \cosh 2\alpha ~G^{(1)}_0(Z) + 
\sinh 2\alpha~ \left[ G_0^{(1)}(-Z) \right]~,\\
D_\alpha (x,y) \!\!\! &= &\!\!\! D_{0}(x,y)~.
\eea
If we decompose $\widetilde{G^F_\alpha}$ similarly, we find that the Hadamard function is
the same as for Allen's propagator, $\widetilde{G_\alpha^{(1)}}=G_\alpha^{(1)}$, but the
antisymmetric part is different and given by
\beq\label{commutator}
i{\rm Sgn}(x,y)D_0(x,y)\cosh2\alpha +
i{\rm Sgn}(x,\overline{y})D_0(x,\overline{y})\sinh2\alpha~,
\eeq
as expected from \eq{kg2}.  Since the imaginary part of the two-point function reflects
the production of particles ``on-mass-shell", \ie on classical geodesics, we would expect
it to reflect particle creation at both points as is evident in \eq{commutator}.  This
illustrates an important way in which the analytic structure of correlation functions
differs in MA-vacua and in the corresponding alpha states.

For simplicity we have written our expressions in this section for  the case of mode-independent 
$\alpha_n$. This could certainly be relaxed, but the equations would become
more cumbersome. It is clear however, that the basic conclusions are independent of this
idealization. For example, in the spirit of effective field theory, we could consider
constant $\alpha_n$ below some large cutoff, and vanishing $\alpha_n$ above the cut-off.
Then the various sources would be smeared; but the conclusion would remain that there are
sources for excited states also at the antipodal points, albeit smeared ones.

\section{The Interacting Theory}
\label{sect:feynman}
We now turn out attention to the interacting theory. 
As in {\bf I}, the Feynman rules for perturbation theory may be 
obtained from 
\beq\label{rules}
\exp\{iW[J]\}=\exp\{i\int dx\sqrt{g} \lcal_I(-i\frac{\delta}{\delta
J})\}\exp\{iW_f[J]\}~,
\eeq
where $W[J]$ is the generating functional of connected Green's functions, and
$\lcal_I(\phi)$ is the interaction Lagrangian density (assumed in this formula to be
nonderivative).  $W_f[J]$ is the free field generating functional given
by\footnote{Although we write the integral in terms of coordinates, we mean the
coordinate-independent integration over the entire de~Sitter manifold. If global
coordinates are chosen, this is manifest. Otherwise, the integral must be defined by
integration over various coordinate patches. The existence of horizons for certain
coordinate systems complicates the discussion, but they do not present any problems of
principle.  One need only replace them by other coordinates in the neighborhood of such
horizons; de~Sitter space is everywhere nonsingular.}
\beq\label{freeW}
W_f[J]=\half\int dx \sqrt{g(x)} dy \sqrt{g(y)}J(x) G^F(x,y) J(y)~.
\eeq
Although we shall assume without proof that the interacting theory in the Euclidean vacuum
is well-defined, at least perturbatively, some comments may be in order.  The reasons for
our confidence in this vacuum are essentially the same as in Minkowski space. With the use
of the Euclidean Feynman propagator in \eq{freeW}, correlation functions may be defined by
Wick rotation from Euclidean signature, and, correspondingly, the Feynman rules yield
amplitudes whose integrands are singularity-free for Euclidean signature.  As a result,
the usual apparatus of perturbation theory goes through.  The derivation of \eq{rules} is
especially straightforward in the path integral formalism, but it can also be performed in
the operator formalism. One may pass from the Heisenberg picture to the interaction
picture and develop the analogues of the Gell-Mann-Low formula and the Dyson expansion for
vacuum expectation values (VEVs) of Green's functions.  As a result, the counterterms
needed to renormalize the field theory in  the Euclidean vacuum are local.  Further, the
K\"allen-Lehmann spectral representation of the two-point functions~\cite{Bros} may then
be extended from the free to the interacting theory, \beq\label{KLrep} G(x,y)=\int d\sigma
\rho(\sigma) G(x,y;\sigma)~, \eeq where $G(x,y;\sigma)$ is the free field two-point
function for a particle of mass-squared $\sigma.$

In {\bf I} it was argued that the analytic structure of the propagator \eq{MAprop} renders
an interacting field theory in an $\alpha$-vacuum ill-defined. The thesis of the present
work is to argue that, in contrast, interactions can be included if the states
$\ket{\underline{\alpha}}$ are regarded as excited states in the Euclidean vacuum, at
least approximately for modes below some cutoff. Then one expects matrix elements of
fields between excited states to be well-defined and calculable using the Feynman rules of
the Euclidean theory.  Moreover, in any sensible formulation, they should be
renormalizable using the same counterterms as for VEVs. The fulfillment of these
expectations is complicated by the fact that the definition of the excited states
corresponding to the free field $\alpha$-states is necessarily more complicated.  The
correct choice will be dictated by the particular physical situation under consideration. 
We shall consider several possibilities.

The first possibility is to use \eq{fpi2} as the definition for correlators in $\alpha$-states 
for an interacting theory just as for a free theory. The computational prescription
is thus to compute interacting correlators in the Euclidean theory, and then form the
alpha-state correlators by taking the linear combinations indicated in, {\it e.g.}
\eq{feynman1}, and similarly for higher point functions. It is manifest from this
prescription that the local counterterms of the Euclidean vacuum will suffice for
renormalization. This definition of alpha-states and the corresponding computational
rules are correct when the interaction is adiabatically switched on and off in the distant
past and future. An important example is the conjectured dS/CFT correspondence, in which a kind of
meta-S-matrix~\cite{Witten} is formally introduced, with in- and out-states defined by
reference to global coordinates~\cite{Strominger,SV}. This singles out particular
$\alpha$-states as non-interacting asymptotic states on $\ical^+$ and $\ical^-.$ For an
interacting field theory, one would have to use the definition of alpha-states discussed
here, in order to have a well-defined field theory; so the in- and out states of dS/CFT
correspondence should not really be thought of as vacua but as highly-correlated excited
states of the Euclidean theory.\footnote{For even dimensions, one must restore 
the non-CPT-invariant phase factor to the preceding formulae.}  A major drawback 
of defining alpha-states by adiabatically
switching off interactions is that this procedure makes the concept frame dependent. For
example, in planar coordinates common to cosmological applications, the distant past 
of a particular observer is the light-cone of an apparent horizon.  

A second possibility for defining the alpha-states is to apply free field definitions
such as \eqs{alphavac}{alphastate} directly in the interacting theory. This is similar to
the definition used, for example, by Danielsson~\cite{udan}, although interactions were
incidental to that work. The problem with this procedure is that the creation and
annihilation operators are time-dependent.  At best, then, one might employ these
equations at some fixed time or, more generally, on some Cauchy surface.  One may
canonically quantize the theory on such a spacelike section and then interpret the system
as being in such a state at that time. In that case, the transformation between $\phi$ and
$\widetilde{\phi},$ \eqs{phitilde}{phitilde2}, must be interpreted at that time, and one
must solve for the behavior of correlators at other times. Although well-defined in
principle, it generally seems intractable to carry out this procedure in practice.
However, for two-point functions, analyticity and de~Sitter invariance are sufficient to
go from correlation functions at equal times to two arbitrary spacetime points, using the
K\"allen-Lehman representation \eq{KLrep}.  Stated otherwise, knowing the two-point
function for all points on a spacelike surface determines it for all times.  Thus, for the
two-point correlators the prescription again becomes taking linear combinations, as in the
free theory~\eqs{wightman}{feynman2}. For applications such as the density fluctuations in
the cosmic microwave background, it is in fact the two-point functions that are of primary
interest so this prescription could perhaps be used to justify the sorts of calculations
in refs.~\cite{udan,transplanck}. However, it is important to note that it is not just the
short-distance modifications that distinguish our interpretation from some of those, but
rather the long-distance, on-shell structure of the states. In our framework, these states
have nothing to do with ``trans-Planckian" physics.

There may be other definitions of alpha-states at the interacting level, appropriate in
other applications. For example, in planar coordinates, commonly used in discussion of
inflation, it is common to speak of in-states defined along the null-surface at conformal
time in the distant past, and one may define $\alpha$-in states there. For measurements
involving Unruh detectors, which refer to in-in correlation functions rather than in-out
correlators, these would be the relevant states to consider. Presumably one may develop a
formalism for evaluating such correlators similar to the real-time formalism in 
finite-temperature perturbation theory~\cite{lebellac}.

Whichever definition is used for alpha-states, the calculation of Green's functions for
excited states involves only the Feynman rules and the counterterms of the Euclidean field
theory.  No non-local counterterms are required.\footnote{As noted previously, we disagree
with~\cite{Goldstein:2003ut} in this respect. Compare also ref.~\cite{Banks}}  This would
be manifest in the determination of the spectral density in the K\"allen-Lehmann
representation for the renormalized field.

\section{Applications and Discussion}
\label{sect:apps}
Our discussion has interesting implications for the Hadamard
condition~\cite{Hadamard,Fulling:nb}, a test often imposed to determine an
acceptable vacuum in a curved space setting.  The Hadamard theorem requires,
among other things, that the leading short distance singularity in the Hadamard
function $G^{(1)}$ should take its flat space value.  As noted in Section 3, the
leading singularity of the Hadamard function is $\cosh 2\alpha$ times its flat
space value, for the alpha-vacua as well as the alpha-states. However, there are
several reasons why this is not adequate to reject alpha-vacua out of hand. We
have already pointed out that in order to ensure that their overlap with the
Euclidean Fock space, MA-states must be cut off at some high scale, above which
it might asymptote to the Euclidean vacuum. This makes the discussion of the
singularity structure at short distances subtle, to say the least.  Such cut-off
states will formally satisfy the Hadamard condition.  The problem is that, in
the limit that the cutoff is removed, they do not. To our mind, that does not
mean that there is no sensible low-energy physics associated with such states.
Indeed, the main theme of this paper has been that alpha-vacua are acceptable if
interpreted as excited states, but not if they are treated as vacua.   When
comparing the Hadamard function for the MA-vacua with that of the excited state
(see \eq{newgfcts} {\it et seq.}\/), we found they were the same.  So the Hadamard
function does not discriminate between the two situations.  

This conundrum is further aggravated by interactions, 
since then the quantum field suffers wave
function renormalization with the consequence that the bare field satisfies
canonical commutation relations (CCR) but must be cutoff, and the renormalized
field does not satisfy CCR but has cutoff-independent correlation functions. In
our view the ingredient needed to improve this situation is the analytic
properties of correlation functions, a crucial tool in Minkowski space that has been insufficiently
exploited heretofore in curved spacetime. We are encouraged in this program that one can distinguish
on theoretical grounds between the treatment of the alpha-vacua as no-particle
states and their interpretation as excited states of the unique Euclidean
vacuum.  

There is no S-matrix in de~Sitter space, and we have not addressed the important issue of
what are observables or how to relate the $n$-point functions to them.    This question is
not peculiar to considering excited states and is not the focus of this paper. We assume
that whatever they are, it is sufficient to know how to calculate the $n$-point functions
for VEVs. At the very least, one must entertain successive measurements by idealized Unruh
detectors, as was assumed in {\bf I}. This suggests that one should consider in-in matrix
elements of fields, \eg, as with the two-point response function.  The rules for relating
in-in matrix elements to Wightman functions are more complicated than for S-matrix
elements, but presumably can be extended from vacuum amplitudes to alpha-states also using
the methods given in this paper.

As discussed in Section 3, in an alpha-state the response to a source at the point $x$ is
particle production both locally at the point $x$ and nonlocally at the antipodal point
$\overline{x}.$ The latter sounds highly acausal and impermissible in a sensible theory.
However, the situation is very much analogous to that seen in gedanken experiments of the
EPR type~\cite{EPR}. It is indeed possible to have nonlocal, seemingly acausal effects, in
the presense of highly correlated states. Of course it is contrary to normal experience to
entertain such highly correlated states, because interactions almost certainly wash out
such correlations.  In cosmology, however, gravitational effects are out of equilibrium,
and the inflaton field in particular is especially weakly interacting. If one is willing
to imagine  that the state of the universe has such correlations over large distances
built in from the beginning, by assumption or design, then it is possible that they can be
maintained until they cross the cosmological horizon and freeze out. Then the question becomes
whether such highly correlated initial states represent physically acceptable or
attractive alternatives for the approximate initial state just prior to the onset of
inflation. Some have argued that such a situation can arise naturally in certain kinds of
hybrid inflation models~\cite{Burgess} whereas others note that such states are almost
impossible to\ generate and maintain~\cite{Kaloper:2002cs}. It certainly seems bizarre to
imagine that such correlations were built in; but so much about our present understanding
of the big bang seems so highly contrived that the supposition that there is such a degree
of coherence would not seem to be ruled out. Of course, there are no particular reasons to
prefer these squeezed states over other possible excited states.  It is a matter of the
pre-Big Bang physics and their consequences for the inflationary paradigm.

\section*{Acknowledgements}
We thank V.~Balasubramanian, T.~Banks, C. Burgess, R.~Holman, N.~Kaloper, and D.~Lowe for discussions. This work was supported in part by DoE Division of High Energy Physics.


\end{document}